\documentclass[manuscript]{aastex}

\usepackage[outdir=./]{epstopdf}
\usepackage{graphics,graphicx}
\usepackage{appendix}
\usepackage{xcolor}

\DeclareUnicodeCharacter{0301}{\'{e}}

\def\arcsa#1#2{$#1^{\prime\prime}_{^\textrm{.}}#2$}

\begin{document}

\title{ALMA Survey of Orion Planck Galactic Cold Clumps (ALMASOP): Deriving Inclination Angle and Velocity of the Protostellar Jets from their SiO Knots
}


\author{
Kai-Syun Jhan\altaffilmark{1,2}, Chin-Fei Lee\altaffilmark{2,1}, Doug Johnstone\altaffilmark{3,4}, Tie Liu\altaffilmark{5}, Sheng-Yuan Liu\altaffilmark{2}, Naomi Hirano\altaffilmark{2}, Ken’ichi Tatematsu\altaffilmark{6,7}, Somnath Dutta\altaffilmark{2}, Anthony Moraghan\altaffilmark{2}, Hsien Shang\altaffilmark{2}, Jeong-Eun Lee\altaffilmark{8}, Shanghuo Li\altaffilmark{9}, Chun-Fan Liu\altaffilmark{2}, Shih-Ying Hsu\altaffilmark{10,2}, Woojin Kwon\altaffilmark{11,12}, Dipen Sahu\altaffilmark{2}, Xun-Chuan Liu\altaffilmark{5}, Kee-Tae Kim\altaffilmark{9,13}, Qiuyi Luo\altaffilmark{5}, Sheng-Li Qin\altaffilmark{14}, Patricio Sanhueza\altaffilmark{15,16}, Leonardo Bronfman\altaffilmark{17}, Zhang Qizhou\altaffilmark{18}, David Eden\altaffilmark{19}, Alessio Traficante\altaffilmark{20}, and Chang Won Lee\altaffilmark{9,13} \\
ALMASOP Team
}

\affil{$^{1}$Graduate Institute of Astronomy and Astrophysics, National Taiwan University, No. 1, Sec. 4, Roosevelt Road, Taipei 10617, Taiwan}

\affil{$^{2}$Academia Sinica Institute of Astronomy and Astrophysics, No. 1, Sec. 4, Roosevelt Road, Taipei 10617, Taiwan}

\affil{$^{3}$National Research Council of Canada, Herzberg, Astronomy and Astrophysics Research Centre, 5071 West Saanich Road, V9E 2E7 Victoria (BC), Canada
}

\affil{$^{4}$Department of Physics and Astronomy, University of Victoria, Victoria, BC V8P 5C2, Canada}

\affil{$^{5}$Shanghai Astronomical Observatory, Chinese Academy of Sciences, 80 Nandan Road, Shanghai 200030, People's Republic of China}

\affil{$^{6}$Nobeyama Radio Observatory, National Astronomical Observatory of Japan, National Institutes of Natural Sciences, 462-2 Nobeyama, Minamimaki, Minamisaku, Nagano 384-1305, Japan}

\affil{$^{7}$Department of Astronomical Science, SOKENDAI (The Graduate University for Advanced Studies), 2-21-1 Osawa, Mitaka, Tokyo 181-8588, Japan}

\affil{$^{8}$School of Space Research, Kyung Hee University, Yongin-Si, Gyeonggi-Do 17104, Republic of Korea}

\affil{$^{9}$Korea Astronomy and Space Science Institute (KASI), 776 Daedeokdae-ro, Yuseong-gu, Daejeon 34055, Republic of Korea}

\affil{$^{10}$Graduate Institute of Physics, National Taiwan University, No. 1, Sec. 4, Roosevelt Road, Taipei 10617, Taiwan}

\affil{$^{11}$Department of Earth Science Education, Seoul National University, 1 Gwanak-ro, Gwanak-gu, Seoul 08826, Republic of Korea}

\affil{$^{12}$ SNU Astronomy Research Center, Seoul National University, 1 Gwanak-ro, Gwanak-gu, Seoul 08826, Republic of Korea}

\affil{$^{13}$University of Science and Technology, Korea (UST), 217 Gajeong-ro, Yuseong-gu, Daejeon 34113, Republic of Korea}

\affil{$^{14}$Department of Astronomy, Yunnan University, and Key Laboratory of Particle Astrophysics of Yunnan Province, Kunming, 650091, People’s Republic of China}

\affil{$^{15}$National Astronomical Observatory of Japan, National Institutes of Natural Sciences, 2-21-1 Osawa, Mitaka, Tokyo 181-8588, Japan}

\affil{$^{16}$Department of Astronomical Science, SOKENDAI (The Graduate University for Advanced Studies), 2-21-1 Osawa, Mitaka, Tokyo 181-8588, Japan}

\affil{$^{17}$Departamento de Astronomía, Universidad de Chile, Casilla 36-D, Santiago, Chile}

\affil{$^{18}$Center for Astrophysics | Harvard \& Smithsonian, 60 Garden Street, Cambridge, MA 02138, USA}

\affil{$^{19}$Armagh Observatory and Planetarium, College Hill, Armagh, BT61 9DB, United Kingdom}

\affil{$^{20}$IAPS-INAF, via Fosso del Cavaliere 100, I-00133, Rome, Italy}

\altaffiltext{}{E-mail: ksjhan@asiaa.sinica.edu.tw, cflee@asiaa.sinica.edu.tw}

\begin{abstract}

We have selected six sources (G209.55-19.68S2, G205.46-14.56S1$_{-}$A,
G203.21-11.20W2, G191.90-11.21S, G205.46-14.56S3, and G206.93-16.61W2) from
the Atacama Large Millimeter/submillimeter Array Survey of Orion Planck
Galactic Cold Clumps (ALMASOP), in which these sources have been mapped in
the CO (J=2-1), SiO (J=5-4), and C$^{18}$O (J=2-1) lines.  These sources
have high-velocity SiO jets surrounded by low-velocity CO outflows. The SiO
jets consist of a chain of knots. These knots have been thought to be
produced by semi-periodical variations in jet velocity.  Therefore, we adopt
a shock-forming model, which uses such variations to estimate the
inclination angle and velocity of the jets.  We also derive the inclination
angle of the CO outflows using the wide-angle wind-driven shell model, and
find it to be broadly consistent with that of the associated SiO jets.  In
addition, we apply this shock-forming model to another three protostellar
sources with SiO jets in the literature -- HH 211, HH 212, and L1448C(N) --
and find that their inclination angle and jet velocity are consistent with
those previously estimated from proper motion and radial velocity studies.

\end{abstract}


\section{Introduction}


Protostellar jets are important components in star formation. They are
launched from the innermost parts of accretion disks and carry away angular
momentum, allowing the disk material to fall onto the protostars
\citep{2014prpl.conf..451F,2020A&ARv..28....1L}.  However, the location
where this launching process occurs ($<$1 au) is too close to the
protostar to be spatially resolved with current observational facilities. 
Therefore, protostellar jet properties, such as velocity and mass-loss rate,
are used to constrain the launching mechanism.  However, it requires
multi-epoch observations with enough spatial and velocity resolution over
several years to measure the jet proper motion in order to derive the jet
velocity.  In addition, inclination angle of the jet, which is also needed
to derive the jet velocity, can not be measured directly.  Here we employ a
novel method to constrain both the inclination angle and the jet velocity,
requiring only one epoch of observation.


Now with Atacama Large Millimeter/submillimeter Array (ALMA), protostellar
jets can be resolved with sufficient resolution.  They have knotty (shock)
structures traced by shock tracers, e.g., SiO
\citep{1998A&A...333..287G,2006ApJ...636L.141H,2006ApJ...636L.137P,2007A&A...462L..53C,2007ApJ...659..499L,Podio2021},
H$_{2}$ \citep{1994ApJ...436L.189M,1998Natur.394..862Z}, and SO
\citep{2007ApJ...659..499L,2010ApJ...713..731L,Podio2021}. These knots have been thought to trace the (internal) shocks produced by
semi-periodical variations in jet velocity at the launching points \cite[see,
e.g.,][]{1990ApJ...364..601R,1993ApJ...413..210S,1997A&A...318..595S,2001ApJ...557..429L}.
 Therefore, we adopt the shock-forming model
\citep{1990ApJ...364..601R,2004ApJ...606..483L}, which uses such variations,
to derive the inclination angle and jet velocity from the knots in the
jets. This model has also been used to account for the lack of SiO knotty shocks
near the central source in HH 211 \citep{2021ApJ...909...11J}. 

In this Letter, we report the detections of protostellar jets and
outflows in the SiO (J=5-4) and CO (J=2-1) transitions with ALMA in six
Class 0 and I sources, G209.55-19.68S2 (HOPS 10), G205.46-14.56S1$\_$A (HOPS
358), G203.21-11.20W2, G191.90-11.21S, G205.46-14.56S3 (HOPS 315) and
G206.93-16.61W2 (HOPS 399) (hereafter G209, G205, G203, G191, G205S3, and
G206 respectively), and test the shock-forming model.  We present the
observations in Section 2.  In Section 3, we present the results of the jets
and use the shock-forming model to estimate the inclination angle and
velocity of the jets.  We also present the results of the outflows and use
the wind-driven shell model \citep{2000ApJ...542..925L} to derive the
inclination angle of the outflows.  In Section 4, we compare the inclination
angles derived from both models, apply the shock-forming model to other
sources reported in the literature, and then discuss jet and outflow
properties, and jet knot timescales.  We summarize our conclusions in
Section 5.


\section{Observations}

Observations toward the G209, G205, G203, G191, G205S3, and G206 systems
were obtained with ALMA (ID: 2018.1.00302.S; PI: Tie Liu). The data are part of the ALMA Survey of Orion
Planck Galactic Cold Clumps (ALMASOP) project
\citep{2020ApJ...898..107H,2020ApJS..251...20D,2021ApJ...907L..15S},
in which only these six targeted sources were seen with clear
periodical knotty structures.
We have mapped these targeted sources in C$^{18}$O J=2-1
(219.560 GHz) to determine their systemic velocities, and mapped their
outflows and jets in the CO J=2-1 (230.538 GHz) and SiO J=5-4 (217.105 GHz)
lines.  More details on the calibration and the line observations can be
seen in \citet{2020ApJS..251...20D} and \citet{2022ApJ...925...11D}.  Here
we only summarize the important parameters for the three lines.  The
CASA 5.4 package was used to image the data.  These sources were observed
with three different configurations, resulting in three different data sets
(TM1, TM2, and ACA).  After combining these data sets, we used the TCLEAN
task with a robust weighting factor of 0.5 to generate various
line-intensity maps with a velocity resolution of $\sim$ 1.6 km s$^{-1}$,
without primary beam correction.  The synthesized beam has a size of $\sim$
\arcsa{0}{35} $\times$ \arcsa{0}{31} for CO, $\sim$ \arcsa{0}{40} $\times$
\arcsa{0}{32} for SiO, and $\sim$ \arcsa{0}{43} $\times$ \arcsa{0}{31} for
C$^{18}$O.  The resulting CO, SiO, and C$^{18}$O channel maps have a noise
level of $\sim$ 2.2, 2.0, and 2.7 mJy beam$^{-1}$, respectively.

\section{Results and models}

In each source, the systemic velocity $V_{sys}$ is assumed to be the
velocity at the peak emission of the C$^{18}$O position-velocity (PV)
structure cut perpendicular to the jet axis going through the sources'
center.  It is found to be 8.0$\pm$0.8, 9.6$\pm$0.8, 9.6$\pm$0.8,
10.4$\pm$0.8, 9.9$\pm$0.8, and 8.8$\pm$0.8 km s$^{-1}$ for G209, G205, G203,
G191, G205S3, and G206, respectively \citep[see Figure \ref{c18o} in
Appendix and Figure 4 in][]{2022ApJ...925...11D}.  Throughout this paper,
the velocity of jet or outflow is the velocity offset relative to the
systemic value.

\subsection{SiO jets and the shock-forming model}
\label{sec-model}

\begin{figure}
\epsscale{0.05}
\includegraphics[scale=0.7,angle=270]{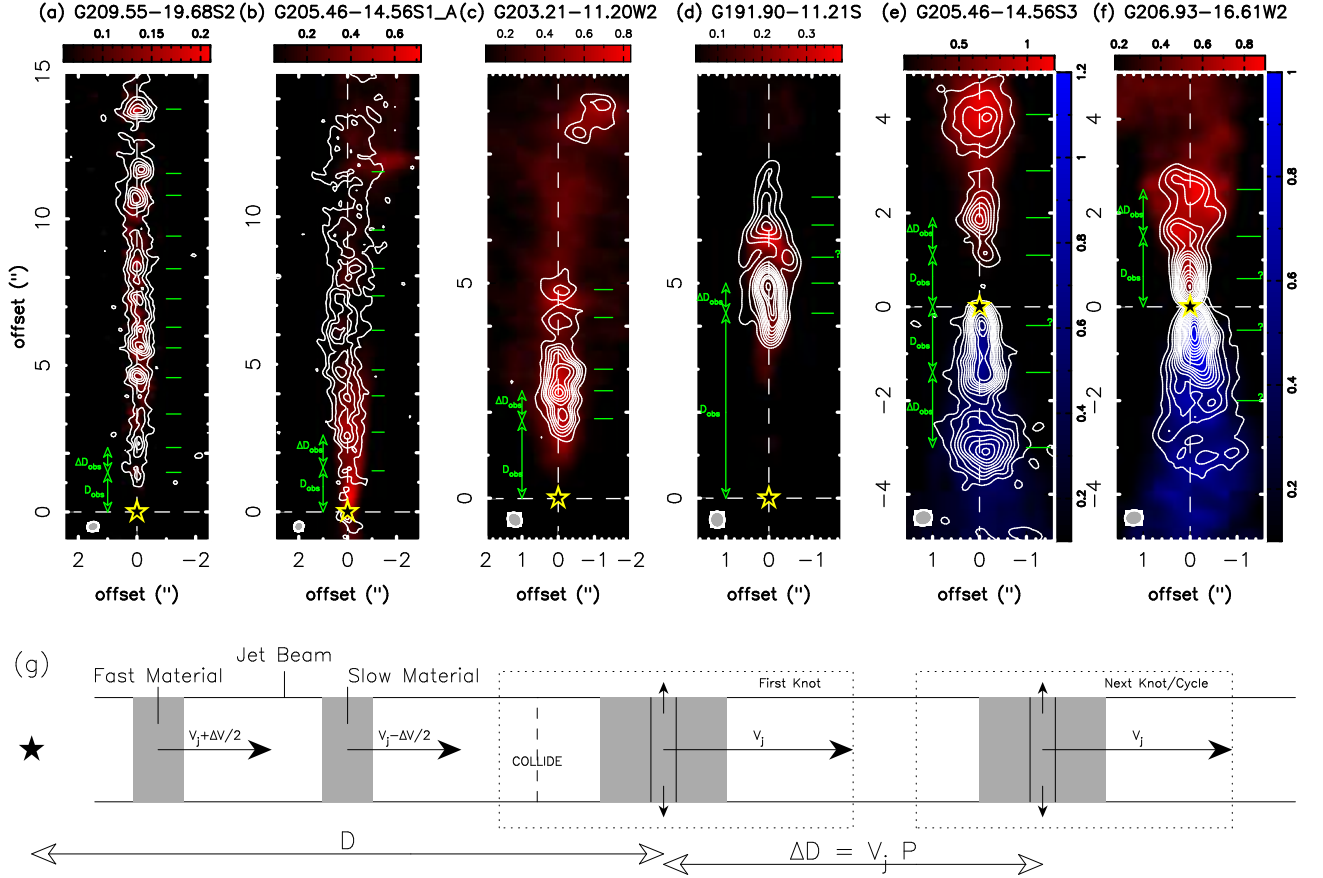}
\centering
\caption{ Panel (a) to (f): CO (red color for redshifted and blue color for
blueshifted) intensity maps and SiO (white contour) intensity maps for the six ALMASOP
sources, G209 (HOPS 10), G205 (HOPS 358), G203, G191, G205S3 (HOPS 315), and
G206 (HOPS 399).  These maps have been rotated clockwise by 29$^{\circ}$,
158$^{\circ}$, 146.7$^{\circ}$, -5$^{\circ}$, -135$^{\circ}$, and -20$^{\circ}$,
respectively, to align the jets with the vertical axis.  For SiO, the contour
levels start at 3$\sigma$ with a step of 3$\sigma$, and $\sigma$ are
$\sim$ 0.02, 0.04, 0.01, 0.02, 0.03, and 0.06 Jy beam$^{-1}$ km s$^{-1}$ in
panel (a) to (f), respectively.  The green line segments indicate the SiO
peak positions of the knots.  The yellow-black stars denote the central source
position.  The green arrows mark the lengths of $D_{\rm obs}$ and $\Delta
D_{\rm obs}$.  Panel (g): Schematic diagram showing how fast and slow jet
material interact to form internal shocks seen as knots
\citep{1990ApJ...364..601R,1993ApJ...413..210S,1997A&A...318..595S,2004ApJ...606..483L,2016ApJ...816...32J}.
\label{fig:SiOjet}
}
\label{jet}
\end{figure}

Figures \ref{jet} (a) to (f) show the SiO maps of our six sources.  For all
targets the SiO emission (contours) is highly collimated and consists of a
chain of knots tracing the jets.  These jets can also be traced by
high-velocity CO gas, as shown in color images. The underlying jets
could be continuous structures, and SiO mainly traces the internal shocks,
where the density is high.  Therefore, throughout this Letter, “SiO jet”
means a jet-like structure consisting of a chain of high-density shocked
regions detected in SiO.
 Notice that four of the
jets (G209, G205, G203, and G191) happen to be monopolar jets in both SiO
and CO and are seen only on the redshifted side, and the other two jets
(G205S3 and G206) are bipolar.

Interestingly, there is no SiO emission between the central sources and
their first (closest) knots in four of the jets.  This phenomenon has been
observed in other sources, e.g., HH 211 \citep{2021ApJ...909...11J}, and can
be explained by a shock-forming model.  In this model, the
sources eject jet material with a periodical variation in the jet velocity
to form internal shocks
\citep[knots,][]{1990ApJ...364..601R,1993ApJ...413..210S,1997A&A...318..595S,2004ApJ...606..483L},
producing SiO in the gas phase and thus the SiO emission
\citep{1997A&A...321..293S}.  However, an internal shock does not
form immediately near the source because it takes time (or distance) for the
fast material to catch up with the slow material, as shown in Figure \ref{jet} (g).



In the model, the jet is assumed to have the following velocity,
\begin{equation}
V = V_{\rm j} - \frac{\Delta V}{2} \sin \left(\frac{2\pi t}{P}\right),
\end{equation}
where $V$ is the jet velocity at a certain position, $V_{\rm j}$ is the mean jet velocity, $\Delta V$ is the amplitude of
the velocity variation, $t$ is the time, and $P$ is the period of the variation \citep{1990ApJ...364..601R,2004ApJ...606..483L}.

An internal shock forms when the fastest material with $V_{\rm j}$ + $\Delta V/2$
catches up (and collides) with the slowest material having $V_{\rm j}$ - $\Delta
V/2$ (see Figure \ref{jet} (g)). The distance to catch-up is given by the
following:

\begin{equation}
D = V_{\rm j} \frac{V_{\rm j}}{\Delta V/(P/\pi)} \,,
\end{equation}
where the value of ($\frac{V_{\rm j}}{\Delta V/(P/\pi)}$) is the time to
catch up \citep{1990ApJ...364..601R,2004ApJ...606..483L}.  After formed, the
shocks are moving at the mean jet velocity (V$_{\rm j}$).  Therefore, the
interval of the knots ($\Delta D$) is given as:

\begin{equation}
\Delta D = PV_{\rm j}.
\end{equation}
Combining Equations 2 and 3 and rearranging, we obtain:


\begin{equation}
V_{\rm j} = \pi \frac{D}{\Delta D} \Delta V.
\end{equation}

\noindent In addition, we assume no energy dissipation so that the 
thermal energy produced by the shocks would be totally transferred sideways
into the sideways ejection. Thus, the sideways-ejection velocity
is equal to $\Delta V$.  Then, the observed radial sideways-ejection
velocity ($\Delta V_{\rm obs}$) and the observed radial jet velocity
($V_{\rm obs}$) are, respectively, given by

\begin{equation}
\Delta V_{\rm obs} = \Delta V\cos i,
\end{equation}
\begin{equation}
V_{\rm obs} = V_{\rm j} \sin i,
\end{equation}
where $i$ is the inclination angle of the jet to the plane of the sky. With Equations 4 and 5, we have:
\begin{equation}
V_{\rm j} \cos i = \pi \frac{D}{\Delta D} \Delta V_{\rm obs} = \pi \frac{D_{\rm obs}}{\Delta D_{\rm obs}} \Delta V_{\rm obs},
\end{equation}
where $D_{\rm obs}$ is the observed (projected to the plane of the sky)
distance from the sources to their first knots and $\Delta D_{\rm obs}$ is
the observed (projected to the plane of the sky) average interval of the
knots.  Here, $\frac{D}{\Delta D}$ is equal to $\frac{D_{\rm obs}}{\Delta
D_{\rm obs}}$ because of the same projection effect.
Then,
from Equations 6 and 7, we can derive the inclination angle with the following
\begin{equation}
i = \tan^{-1} \left(\frac{V_{\rm j} \sin i}{V_{ \rm j} \cos i}\right) = \tan^{-1} \left(\frac{V_{\rm obs}}{\pi \frac{D_{\rm obs}}{\Delta D_{\rm obs}} \Delta V_{\rm obs}}\right).
\end{equation}
using the values measured from the observations, and
then $V_{j}$ with Equation 6.



Table 1 shows the values of $D_{\rm obs}$ and $\Delta D_{\rm obs}$
measured from the total intensity SiO maps (Figure \ref{fig:SiOjet}), the
values of $\Delta V_{\rm obs}$ and $V_{\rm obs}$ measured from the SiO PV
diagrams \citep[Figure \ref{sio} in the Appendix and Figure 1
in][]{2022ApJ...925...11D}, and the corresponding values of $i$, $V_{\rm
j}$, and thus $P$ derived from the shock-forming model.  Here the
distance where the shock first forms $D_{\rm obs}$ is uncertain and is thus
assumed to have an error of $0.5 \Delta D_{\rm obs}$ on the negative side
and a half of the beam size on the positive side, where $\Delta D_{\rm obs}$
is the mean separation between the two consecutive knots.  Notice that the
first knots on the blueshifted side of G205S3 and on the both sides of G206
are too close to the sources to form shocks, and thus should have a different
origin. Therefore, we adopted the knots downstream to be the first knots
formed by shocks.


\subsection{CO outflows and the wide-angle wind-driven shell model}

\begin{figure}
\epsscale{0.5}
\plotone{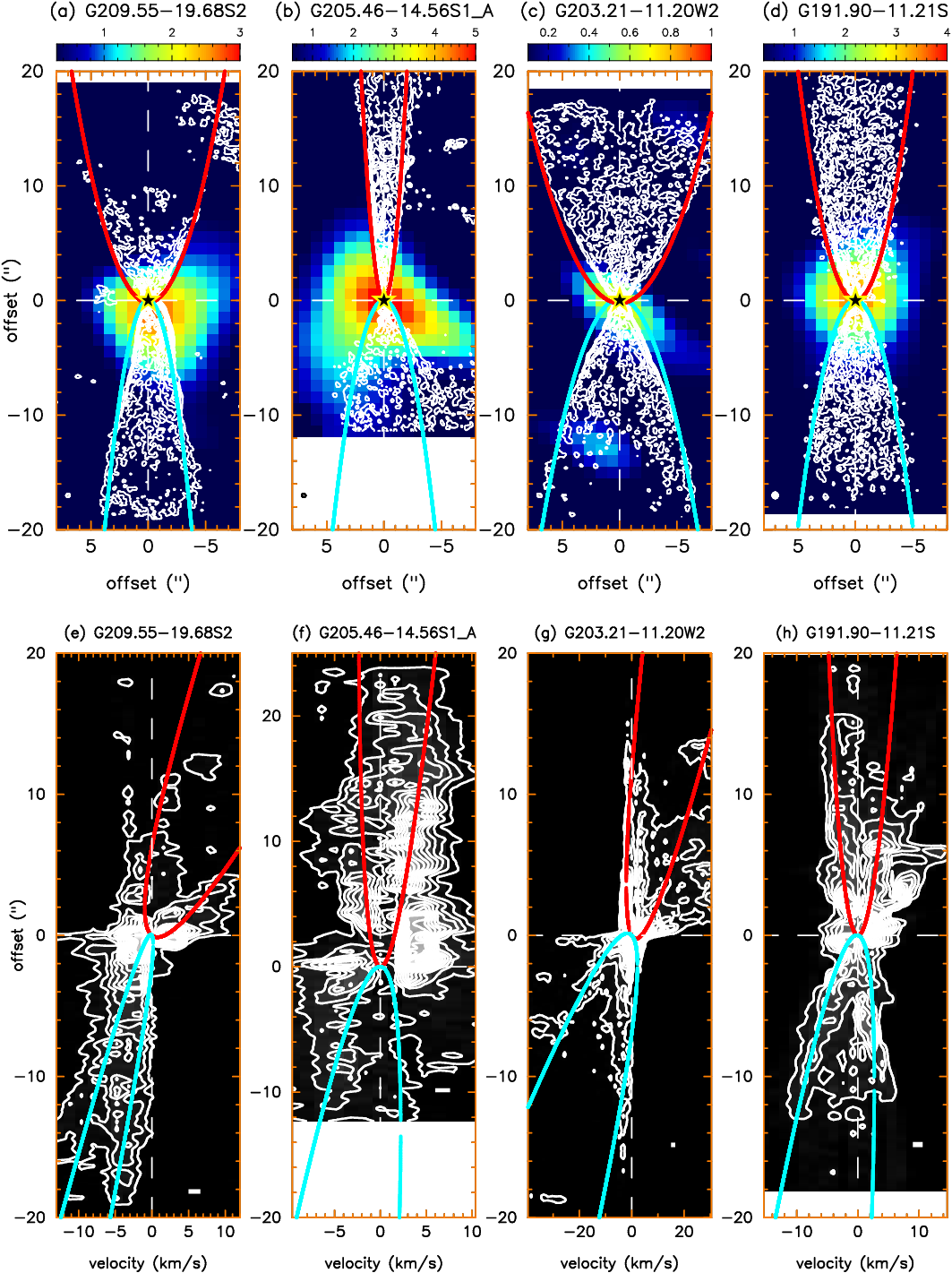}
\centering
\caption{Panel (a) to (d): CO channel maps (contours) and C$^{18}$O 
total intensity maps (color image) for the four ALMASOP sources G209
(HOPS 10), G205 (HOPS 358), G203, and G191.  The contour levels start at
3$\sigma$ with a step of 3$\sigma$, and $\sigma$ are $\sim$0.013, 0.016,
0.013 and 0.016 Jy beam$^{-1}$ km s$^{-1}$ in panel (a) to (d),
respectively.  The yellow-black stars denote the central source position.  Panel
(e) to (h): CO PV diagrams cut along the outflow axis for the four sources.  The
contour levels start at 10$\sigma$ with a step of 10$\sigma$, and $\sigma$
are $\sim$0.002, 0.0012, 0.001 and 0.002 Jy beam$^{-1}$ in panel (e) to
(h), respectively. The blue and red curves are the fits to the
structures and PV structures of the outflows from the wind-driven
shell model (see Section 3.2).
}
\label{outflow}
\end{figure}

CO outflows are detected in blue and red-shifted lobes for all the sources. 
Here we present the CO outflow maps for four of them (Figure
\ref{outflow}), because the CO outflows of G205S3 and G206 have been reported
and analyzed before \citep[Figure 2 (a) and (c) in][]{2022ApJ...925...11D}.
The outflows display shell-like structures around the jets.  Due to the
projection effect, the structure of the outflow shells is better seen at low
velocities close to the systemic velocity, thus we present only one or two
velocity channels close to $V_{\rm sys}$ to show the outflow shell
structures.  As can be seen, these outflow shells have larger opening angles
and thus less collimation than their associated SiO jets.



As discussed in \citet{2000ApJ...542..925L} and \citet{2022ApJ...925...11D},
a wide-angle wind-driven shell model can be used to fit the outflow shells
and obtain their inclination angle, opening angle, and dynamical age.  This
model will thus be used to derive the inclination angle of the outflows to
be compared with that of the jets derived earlier.  In this model, the
outflow shell is assumed to be a radially expanding parabolic shell with its
structure and velocity given in the cylindrical coordinate system $(z,R)$ by
\begin{equation} z = c R^{2} \end{equation} \begin{equation} v_{R} =
\frac{R}{t_{0}}, \end{equation} where $c$ is a free parameter negatively
correlated to the opening angle of the shell structure and $t_{\rm 0}$ is
the dynamical age of the outflow shell to be obtained. The
resulting shell structure projected on the sky $(x^{\prime}, z^{\prime};$ as
horizontal axis and vertical axis in Figure \ref{outflow} (a) to (d) ) and
PV structure along the jet axis $(v_{\rm obs}, z^{\prime}$; as horizontal
axis and vertical axis in Figure \ref{outflow} (e) to (h)) would be:
\begin{equation}
z^{\prime} = cx^{\prime 2} \cos i \left(1 -
\frac{\tan^{2}i}{4c^{2}x^{\prime 2}}\right) 
\end{equation} 
and
\begin{equation} 
z^{\prime} = -\frac{v_{\rm obs}t_{0}}{\tan i} - \frac{1}{2c
\tan i \sin i} \left[-1\pm\sqrt{1 - \frac{4cv_{\rm obs}t_{0} \tan i}{\cos
i}}\right] 
\end{equation}
, respectively, where $i$ is the inclination angle.  
From Equation 11, we can see that the model produces a parabolic
shell with its opening angle mainly depending on $c$.  This parabolic shell
corresponds to the boundary of the observed outflows in the low-velocity
channel maps (Figure \ref{outflow} (a) to (d)).  Based on Equation 12, this
model produces a tilted parabolic PV structure with its tilt and opening
angle depending on $i$, $c$, and $t_{0}$.  Therefore, for each outflow, we
have obtained the values of $i$, $c$, and $t_0$ (see Table \ref{tab:CO}), by
fitting Equation 11 to the boundary of the outflow in the low-velocity
channel maps (Figure \ref{outflow} (a) to (d)), and by searching the tilted
parabolic PV structure and then fitting Equation 12 to its tilt and opening angle
in the PV diagrams (Figure \ref{outflow} (e) to (h)).
Note that the values for G205S3 and G206 are
taken from \citet{2022ApJ...925...11D}.  The errors of these parameters are
estimated from the uncertainty of intensity peak values in the observed
channel maps and PV diagrams (Figure \ref{outflow}).


\section{Discussion}


As can be seen from Table \ref{tab:CO}, the inclination angles
derived for the jets, $i(\rm SiO)$, using the shock-forming model are
broadly consistent (within the errors) with those derived for the associated
outflows, $i(\rm CO)$, using the wind-driven shell model.  The inclination
angles of the jet and the outflow are expected to be roughly the same
because the jet and the underlying wind (that drives the outflow) are
believed to come from the same accretion disk.  Thus, this consistency
supports that the shock-forming model can be used to derive the inclination
angle and velocity of the jets.  Notice that, in order for this model to
work, the jets must have at least a few roughly equally spaced knots
produced by internal shocks.  In addition, the shock speed ($\Delta V$)
should be around 10 to 40 km s$^{-1}$ \citep{1997A&A...321..293S}, so that
the SiO emission is indeed produced by shocks.  Moreover, no significant
energy is dissipated in the shocks so that the thermal energy produced by
the shocks can be totally transferred sideways into the sideways ejection. 


\subsection{Application to other sources}

In order to further test the shock-forming model, we applied it to
three previously reported SiO jets, HH 211
\citep{2016ApJ...816...32J,2021ApJ...909...11J}, HH 212
\citep{2015ApJ...805..186L}, and L1448C(N) \citep{2021ApJ...906..112Y}, of
which the jet velocity and inclination angle have been derived independently
from proper motion measurements and the observed radial velocity. In the
case of HH 211, the first pair of knots (knots BK0 and RK0) are too close ( $<$ 25AU) to
the source, so they are unlikely formed by shocks and thus excluded from our
analysis.  In the case of HH 212, we also included the H$_2$ knots in our
analysis \cite[see Figure 4 and 6 in][]{2015ApJ...805..186L}, because the
shocks there are too strong for SiO to survive and thus are only seen in
H$_2$.  In the case of L1448C(N), the first pair of knots (R1-a and BI-a)
are seen with a large velocity range of $>$ 40 km s$^{-1}$, which is too large a shock velocity for SiO to survive
\citep{1997A&A...321..293S,2000MNRAS.318..809M,2008A&A...482..809G,2013MNRAS.428..381V}. 
Also, their PV structures are not fully resolved and thus
might have a different origin. Therefore, the first pair of
knots in this jet are excluded from our analysis.  As can be seen in Table 1, the derived
inclination angle and velocity of the jet in these three jets are also
broadly consistent with those previously measured from proper motion and
radial velocity.

\subsection{Jets and Outflows}

In our sample, four of the jets detected in SiO are monopolar and are
seen only on the redshifted side, although their associated CO outflows are
bipolar. As mentioned earlier, the same jets are also detected in CO
and also appear to be monopolar on the redshifted side.  This asymmetry of
the jets is unlikely to be caused by an asymmetry of the ambient material,
which can only affect material at low velocity in the CO outflow
\citep{2014A&A...563L...3C}.
 In fact, a monopolar jet has also been detected in SiO in NGC 1333–IRAS2A,
although on the blueshifted side \citep{2014A&A...563L...3C}. Since
dust extinction is negligible in radio wavelength, the lack of a redshifted
jet component in that system is likely intrinsic.  Thus, the authors argued
that the accreting disk in that system is ejecting more material from the
side tilted toward us than the side tilted away from us, producing the
blueshifted monopolar jet.  This scenario can also apply to our jets, but
with the accreting disks ejecting more material from the side tilted away
from us.








The CO outflows also show asymmetry in intensity maps and PV
diagrams (Figure \ref{outflow}), with different opening angles on each side
of the source (column 5 in Table 2).  It could be due to an accumulative
interaction of the jet and wind with the ambient material within their
dynamical age, which is $\sim$ 2000 years (column 6 in Table 2).  Some
asymmetry is seen in the ambient material in the C$^{18}$O total intensity
maps around the sources (Figure \ref{outflow} (a) to (d)), however, further
observations at a larger scale are needed to check it.


In our shock-forming model, the SiO knots in the jets are produced
by a periodical variation in the jet velocity.  The SiO PV structure along
the jet axis (Figure \ref{sio}) also indicates a presence of variation in
the jet velocity, with the fast jet material catching up
with the slow jet material.
The reason for such a variation
is not clear, and it may be caused by an orbital motion of a binary
companion \citep{2008ApJ...676.1088M}, semi-periodical variation of the jet
launching radius \citep{2000prpl.conf..789S,2007prpl.conf..277P}, or the
variation of magnetic field morphology in the star-disk system
\citep{2014prpl.conf..451F}, etc.  Alternatively, the knots could be due to
periodical enhanced mass ejections.  However, it is unclear how such
ejections can form shocks \citep{2014prpl.conf..451F} to produce the
observed SiO knotty shocks.



\subsection{Jet Knot Timescales and Mid-IR/Sub-mm Light Curve Variability}

It is striking that the majority of jets analyzed in this paper have knot timescales on the order of decades (Table \ref{tab-params}), suggesting that there is some recurrent form of disk or accretion instability operating at this frequency. It is therefore useful to compare these jet knot results against long-term, many year, monitoring observations of the mid-IR and sub-mm continuum emission from similar young stellar objects across the Gould Belt.


\citet{park21} analyzed 7-year mid-IR light curves of 735 protostars, classifying 140 (20\%) as having dominant brightness variation timescales longer than three years. Similarly, \citet{lee21} classified 4-year sub-mm light curves of 43 protostars, and recovered 15 (35\%) with dominant periods longer than three years. In both cases, the majority of these light curve variations are longer than the observing window while also showing evidence of curvature, suggesting dominant time scales of decades. Furthermore, the estimated amplitude variations imply order unity changes in the central source luminosity, attributed to order unity variability in the protostellar mass accretion rate \citep{johnstone13}. For sources common to both investigations, the mid-IR and sub-mm light curves of the long-term variables show strong similarities \citep{contreras20, lee21}. 


Three of the nine sources investigated in this paper are monitored in the sub-mm (G205S1, G205S3, and HH 211) and all show long-term variability. A further three sources were included in the mid-IR sample (G205S3, and G209S2, and L144CN) with only L1448C(N) showing definite long-term variability. Separately, \citet{2022ApJ...925...11D} reports that G206W2 is undergoing a long-term mid-IR secular change.  We have checked the mid-IR light curves \citep[NEOWISE,][]{cutri15} for the other five sources by eye. G191S, G205S1, and HH 212 are definite long-term variables. In summary, a clear majority of this paper's jet knot sources show evidence of many-year accretion variability.


\section{Conclusions}


We have studied  the jets and outflows in six sources and mapped them in SiO
(J=5-4) and CO (J=2-1) at high resolution with ALMA.  The jets consist of a
chain of roughly equally spaced SiO knots that can be produced by
semi-periodical variations in jet velocity.  Thus, we have used a
shock-forming model with a periodical variation in jet velocity to estimate
the inclination angle and velocity of the SiO jets in these six sources from
their SiO knots.  We have also used a wide-angle wind-driven shell model to
fit the shell structure and PV diagrams of the CO outflows to derive the
inclination angles of the CO outflows.  The derived inclination angles of
the SiO jets and CO outflows are broadly consistent with each other.  We
also applied the shock-forming model to three additional SiO jets reported
in the literature, and found that the derived jet velocity and inclination
angle are also broadly consistent with those previously estimated from
proper motion and radial velocity.  Our results support that the knots in
the jets are indeed produced by semi-periodical variations in jet velocity,
so that the shock-forming model can be used to determine the velocity and
inclination angle of the jets from the SiO knots. The periods of the
velocity variations are on the order of decades for most of the jets studied
here.

\acknowledgments

We thank the reviewer for constructive comments. This paper makes use of the following ALMA data: ADS/JAO.ALMA\#2018.1.00302.S. ALMA is a partnership of ESO (representing its member states), NSF (USA) and NINS (Japan), together with NRC (Canada), NSC and ASIAA (Taiwan), and KASI (Republic of Korea), in cooperation with the Republic of Chile. The Joint ALMA Observatory is operated by ESO, AUI/NRAO and NAOJ. 
K.-S.J and C.-F.L. acknowledge grants from the Ministry of Science and Technology of Taiwan (MoST 107-2119-M-001-040-MY3, 110-2112-M-001-021-MY3) and the Academia Sinica (Investigator Award AS-IA-108-M01). 
C.W.L. is supported by the Basic Science Research Program through the National Research Foundation of Korea (NRF) funded by the Ministry of Education, Science and Technology (NRF- 2019R1A2C1010851). 
D.J.\ is supported by NRC Canada and by an NSERC Discovery Grant. 
N.H. acknowledges support from MoST 109-2112-M-001-023 and 109-2112-M-001-048 grants.
J.-E. Lee has been supported by the National Research Foundation of Korea (NRF) grant funded by the Korea government (MSIT) (grant number 2021R1A2C1011718). 
P.S. was partially supported by a Grant-in-Aid for Scientific Research (KAKENHI Number 18H01259) of the Japan Society for the Promotion of Science (JSPS). 
LB gratefully acknowledges support by the ANID BASAL projects ACE210002 and FB210003.
S.-L. Qin is supported by the National Natural Science Foundation of China (grant No. 12033005).
Tie Liu acknowledges the supports by National Natural Science Foundation of China (NSFC) through grants No.12073061 and No.12122307, the international partnership program of Chinese Academy of Sciences through grant No.114231KYSB20200009, Shanghai Pujiang Program 20PJ1415500 and the science research grants from the China Manned Space Project with no. CMS-CSST-2021-B06.

\setcounter{figure}{0}
\renewcommand{\thefigure}{A\arabic{figure}}

\appendix

\section{C$^{18}$O PV structure used to derive the systemic velocity}

\begin{figure}
\epsscale{1}
\plotone{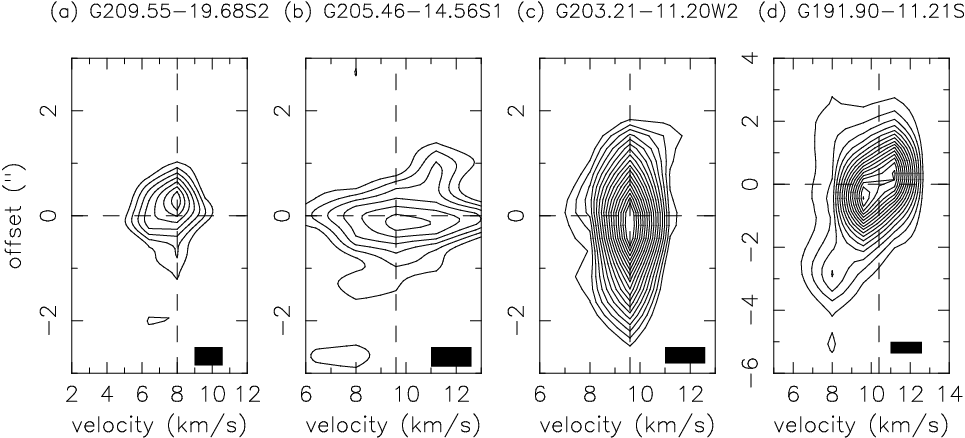}
\centering
\caption{PV diagrams  cut perpendicular to the jet axis going through the sources' center}
for four sources 
(G209, G205, G203, and G191) in C$^{18}$O, centered at the source position. The contour levels start at 3$\sigma$ with a step of 3$\sigma$, and $\sigma$ are $\sim$0.003,
0.003, 0.003 and 0.003 Jy beam$^{-1}$ in panel (a) to (d), respectively.
The vertical dash line indicates the systemic velocity for each source. 
\label{c18o}
\end{figure}

\section{SiO PV structure along the jet axis.}

\setcounter{figure}{0}
\renewcommand{\thefigure}{B\arabic{figure}}

\begin{figure}
\epsscale{0.8}
\plotone{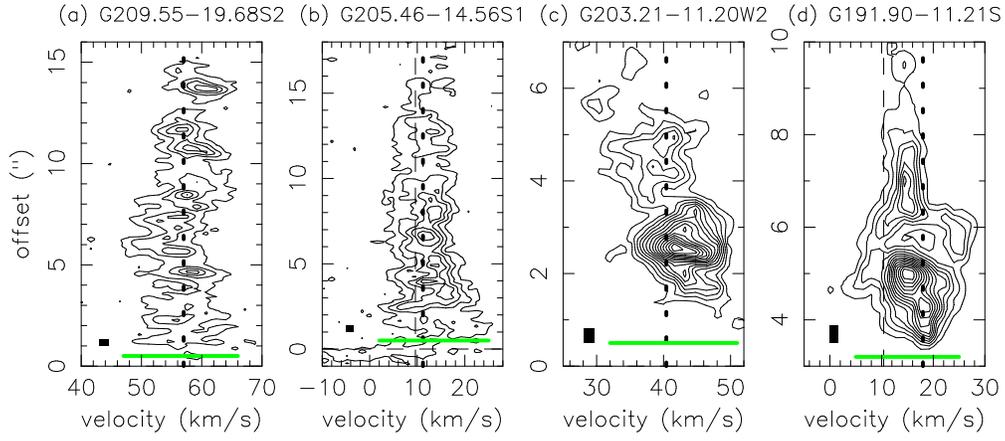}
\centering
\caption{PV diagrams of the jets along the jet axis in SiO. 
The width of the green line segment denotes the 
observed radial sideways-ejection velocity 
($\Delta V_{\rm obs}$) for each source, and 
the dot line denotes the observed radial jet velocity ($V_{\rm obs}$). The contour levels start at 3$\sigma$ with a step of 3$\sigma$, and $\sigma$ are $\sim$0.002,
0.002, 0.002 and 0.0019 Jy beam$^{-1}$ in panel (a) to (d), respectively.
}
\label{sio}
\end{figure}

\begin{table}
\caption{Quantities measured from the SiO jets and the derived
quantities from the shock-forming model}
\label{tab-params}
\begin{tabular}{cccccccccc}
\hline
Source & $D_{\rm obs}$ & $\Delta D_{\rm obs}$ & $\Delta V_{\rm obs}$ & $V_{\rm obs}$ & $i$ & $V_{\rm j}$ & $P$ & Refs.\\
Name & ($\prime\prime$) & ($\prime\prime$) & (km s$^{-1}$)  & (km s$^{-1}$)  & ($^{\circ}$) & (km s$^{-1}$) & (year)\\\hline
G209 (HOPS 10) & 1.3$^{+0.2}_{-0.4}$ & 0.8$\pm$0.2 & 20 & 49 & 28$^{+0.7}_{-0.6}$ & 104$^{+25}_{-20}$ & 17\\\hline

G205 (HOPS 358) & 1.4$^{+0.2}_{-0.5}$ & 1.0$\pm$0.2 & 20 & 1.6 & 1.0$^{+0.6}_{-0.2}$ & 88$^{+12}_{-20}$ & 22\\\hline

G203 & 1.8$^{+0.2}_{-0.5}$ & 1.0$\pm$0.2 & 20 & 30.8 & 18$^{+4}_{-5}$ & 117$^{+12}_{-30}$ & 17 \\\hline

G191 & 4.1$^{+0.2}_{-0.45}$ & 0.9$\pm$0.2 & 20 & 8.9 & 1.8$^{+0.2}_{-0.2}$ & 286$^{+14}_{-30}$ & 6\\\hline

G205S3 (HOPS 315) & 1.1$^{+0.2}_{-0.55}$ & 1.1$\pm$0.2 & 20 & 70 & 48$^{+13}_{-5}$ & 94$^{+8}_{-18}$ & 33\\
& redshifted & & & & & & && \\\hline

G205S3 (HOPS 315) & 1.5$^{+0.2}_{-0.75}$ & 1.5$\pm$0.2 & 20 & 80 & 52$^{+16}_{-4}$ & 102$^{+5}_{-17}$ & 45\\
& blueshifted & & & & & & & & \\\hline

G206 (HOPS 399) & 1.5$^{+0.2}_{-0.5}$ & 1$\pm$0.2 & 40 & 50 & 15$^{+6}_{-2}$ & 195$^{+20}_{-60}$ & 16\\
& redshifted & & & & & & & \\\hline\hline

HH211 & 2.4$^{+0.5}_{-1}$ & 2$\pm$0.5 & 30 & 20 & 10$^{+6}_{-1.7}$ & 114$^{+24}_{-46}$  & 27 & 1 \\\hline

HH211-obs* & 2.6 & 2 & 30 & 20 & 11$\pm$1 & 104 $\pm$16 & &  1 \\\hline

HH212 & 4.5$^{+0.5}_{-1}$ & 2$\pm$0.5 & 20 & 7.3 & 3$^{+0.7}_{-0.3}$ & 141$^{+16}_{-30}$ & 27& 2 \\\hline

HH212-obs* & & & & & 4$\pm$2 & 115$\pm$50 & & 2\\\hline

L1448C(N) & 2.63$^{+0.3}_{-1.1}$ & 2.2$\pm$0.3 & 20 & 54 & 36$^{+0.7}_{-3}$ & 92$^{+7}_{-2}$ & 40 & 3\\
 & redshifted & & & & & & & & \\\hline

L1448C(N) & 1.87$^{+0.3}_{-1.6}$ & 3.2$\pm$0.3 & 20 & 61.2 & 59$^{+30}_{-7}$ & 71$^{+6}_{-10}$ & 122 & 3 \\
 & blueshifted & & & & & & & & \\\hline

L1448C(N)-obs* & & & & & 40$\pm$6 & 88 $\pm$10 & & 3 \\\hline

\end{tabular}
References: (1) \citet{2016ApJ...816...32J,2021ApJ...909...11J} (2) \citet{1998Natur.394..862Z,2015ApJ...805..186L,2017NatAs...1E.152L} (3) \citet{2010ApJ...717...58H,2021ApJ...906..112Y}

*from proper motion measurements and their radial velocity.
\end{table}

\begin{table}
\begin{center}
\caption{Inclination angle of the SiO jets and
the derived quantities of the CO outflows from the wide-angle wind-driven shell model}
\label{tab:CO}
\begin{tabular}{cccccc}
\hline
Source Name & HOPS & $i$ (SiO) & $i$ (CO) & $c$ & $t_{\rm 0}$ \\

 & & ($^{\circ}$) & ($^{\circ}$) & (arcsec$^{-1}$) & (km s$^{-1}$ arcsec$^{-1}$) \\\hline

G209 Red & HOPS 10 & 28$^{+0.7}_{-0.6}$ & 30$\pm$5 & 0.55$\pm$0.05 & 0.7$\pm$0.1\\\hline
G209 Blue & HOPS 10 & & 30$\pm$3 & 1.6$\pm$0.3 &1.3$\pm$0.2\\\hline
G205 Red & HOPS 358 & 1.0$^{+0.6}_{-0.2}$ & 5$\pm$2 & 1$\pm$0.3 & 1.2$\pm$0.3\\\hline
G205 Blue & HOPS 358 & & 4$\pm$1 & 4$\pm$1 & 0.4$\pm$0.15\\\hline
G203 Red & & 18$^{+4}_{-5}$ & 30$\pm$3 & 0.3$\pm$0.1 & 0.6$\pm$0.2\\\hline
G203 Blue & & & 30$\pm$4 & 0.5$\pm$0.1 & 0.35$\pm$0.05\\\hline
G191 Red & & 1.8$^{+0.2}_{-0.2}$ & 2$\pm$0.5 & 0.8$\pm$0.2 & 0.9$\pm$0.2\\\hline
G191 Blue & & & 10$\pm$2 & 0.8$\pm$0.1 & 0.65$\pm$0.1\\\hline
G205S3 Red & HOPS 315 & 48$^{+13}_{-5}$ & 40$\pm$8$^{*}$ & 0.26$^{*}$ & \\\hline
G205S3 Blue & HOPS 315 & 52$^{+16}_{-4}$ & 40$\pm$8$^{*}$ & 0.2$^{*}$ & \\\hline
G206 Red & HOPS 399 & 15$^{+6}_{-2}$ & 10$\pm$5$^{*}$ & 0.8$^{*}$ & \\\hline

\end{tabular}
\end{center}
NOTE - $^{*}$ Reference: \citet{2022ApJ...925...11D}

\end{table}


\end{document}